\documentclass[
superscriptaddress,
twocolumn,
 amsmath,amssymb,
 aps,
pra,
]{revtex4-1}

\usepackage{xcolor}
\usepackage{float}
\usepackage{comment}

\usepackage{graphicx}
\usepackage{dcolumn}
\usepackage{bm}

\usepackage[english]{babel}
\usepackage{float}
\usepackage{graphicx} 
\usepackage[export]{adjustbox}

\begin{document}

\title{Dissipative dynamics of atomic and molecular Rydberg gases:  Avalanche to ultracold plasma states of strong coupling}

\author{M. Aghigh} 
\affiliation{Department of Chemistry, University of British Columbia, Vancouver, BC V6T 1Z3, Canada}
\author{K. Grant} 
\affiliation{Department of Chemistry, University of British Columbia, Vancouver, BC V6T 1Z3, Canada}
\author{R. Haenel}
 \affiliation{Department of Physics \& Astronomy, University of British Columbia, Vancouver, BC V6T 1Z3, Canada}
\author{K.  L. Marroqu\'in} 
\affiliation{Department of Chemistry, University of British Columbia, Vancouver, BC V6T 1Z3, Canada}
\author{F. B. V. Martins} 
\altaffiliation[Present address: ] {Department of Chemistry and Applied Biosciences, ETH Zurich, 8093 Zurich, Switzerland}
 \affiliation{Department of Chemistry, University of British Columbia, Vancouver, BC V6T 1Z3, Canada}
\author{H. Sadegi} 
\altaffiliation[Present address: ] {D-Wave Systems Inc., 3033 Beta avenue, Burnaby BC, V5G4M9, Canada}
 \affiliation{Department of Chemistry, University of British Columbia, Vancouver, BC V6T 1Z3, Canada}
\author{ M. Schulz-Weiling}
\altaffiliation[Present address: ] {Bosch Engineering GmbH, Robert-Bosch-Allee 1, 74232 Abstatt, Germany}
 \affiliation{Department of Physics \& Astronomy, University of British Columbia, Vancouver, BC V6T 1Z3, Canada}
\author{J. Sous}  
\altaffiliation[Present address: ] {Department of Physics, Columbia University, New York, New York 10027, USA}
 \affiliation{Department of Physics \& Astronomy, University of British Columbia, Vancouver, BC V6T 1Z3, Canada}
\author{R. Wang}  
\affiliation{Department of Physics \& Astronomy, University of British Columbia, Vancouver, BC V6T 1Z3, Canada}
\author{J. S. Keller}
 \altaffiliation[Permanent address: ] {Department of Chemistry, Kenyon College, Gambier, Ohio 43022 USA}
 \affiliation{Department of Chemistry, University of British Columbia, Vancouver, BC V6T 1Z3, Canada}
\author{E. R. Grant}
\email[Author to whom correspondence should be addressed. Electronic mail:  ]
{edgrant@chem.ubc.ca}
\affiliation{Department of Chemistry, University of British Columbia, Vancouver, BC V6T 1Z3, Canada}
\affiliation{Department of Physics \& Astronomy, University of British Columbia, Vancouver, BC V6T 1Z3, Canada}

\begin{abstract}

Not long after metastable xenon was photoionized in a magneto-optical trap, groups in Europe and North America found that similar states of ionized gas evolved spontaneously from state-selected, high principal quantum number Rydberg gases.  Studies of atomic xenon and molecular nitric oxide entrained in a supersonically cooled molecular beam subsequently showed much the same final state evolved from a sequence of prompt Penning ionization and electron-impact avalanche to plasma, well-described by coupled rate-equation simulations.  But, measured over longer times, the molecular ultracold plasma was found to exhibit an anomalous combination of very long lifetime and very low apparent electron temperature.  This review summarizes early developments in the study of ultracold plasmas formed by atomic and molecular Rydberg gases, and then details observations as they combine to characterize properties of the nitric oxide molecular ultracold plasma that appear to call for an explanation beyond the realm of conventional plasma physics.

\end{abstract}

\maketitle


\section{Introduction}

Ultracold neutral plasmas studied in the laboratory offer access to a regime of plasma physics that scales to describe thermodynamic aspects of important high-energy-density systems, including strongly coupled astrophysical plasmas \cite{VanHorn,Burrows}, as well as terrestrial sources of neutrons \cite{Hinton,Ichimaru_fusion,Atzeni,Boozer} and x-ray radiation \cite{Rousse,Esarey}.  Yet, under certain conditions, low-temperature laboratory plasmas evolve with dynamics that are governed by the quantum mechanical properties of their constituent particles, and in some cases by coherence with an external electromagnetic field.   

The relevance of ultracold plasmas to such a broad scope of problems in classical and quantum many-body physics has given rise to a great deal of experimental and theoretical research on these systems since their discovery in the late 90s.  A series of reviews affords a good overview of progress in the last twenty years \cite{Gallagher,Killian_Science,PhysRept,Lyon}.  Here, we focus on the subset of ultracold neutral plasmas that form via kinetic rate processes from state-selected Rydberg gases, and emphasize in particular the distinctive dynamics found in the evolution of molecular ultracold plasmas.  

While molecular beam investigations of threshold photoionization spectroscopy had uncovered relevant effects a few years earlier \cite{Scherzer,Alt}, the field of ultracold plasma physics began in earnest with the 1999 experiment of Rolston and coworkers on metastable xenon atoms cooled in a magneto optical trap (MOT) \cite{Killian}.  

This work and many subsequent efforts tuned the photoionization energy as a means to form a plasma of very low electron temperature built on a strongly coupled cloud of ultracold ions.  Experiment and theory soon established that fast processes associated with disorder-induced heating and longer-time electron-ion collisional rate processes act to elevate the ion temperatures to around one degree Kelvin, and constrain the effective initial electron temperature to a range above 30 K \cite{Kuzmin,Hanson,Laha}.  

This apparent limit on the thermal energy of the electrons can be more universally expressed for an expanding plasma by saying that the electron correlation parameter, $\Gamma_e$, does not exceed 0.25, where, 
\begin{equation}
\Gamma_e = \frac{e^2}{4\pi \epsilon_0 a_{ws}}\frac{1}{k_B T_e}
\label{eqn:gamma_e}
\end{equation}
defines the ratio of the average unscreened electron-electron potential energy to the electron kinetic energy.  $a_{ws}$ is the Wigner-Seitz radius, related to the electron density by, $\rho_e = 1/(\frac{4}{3} \pi a_{ws}^3)$.  These plasmas of weakly coupled electrons and strongly coupled ions have provided an important testing ground for ion transport theory and the study of electron-ion collision physics \cite{Strickler}.

Soon after the initial reports of ultracold plasmas formed by direct photoionization, a parallel effort began with emphasis on the plasma that forms spontaneously by Penning ionization and electron-impact avalanche in a dense ultracold Rydberg gas \cite{Mourachko}.  This process affords less apparent control of the initial electron temperature.  But, pulsed field-ionization measurements soon established that the photoionized plasma and that formed by the avalanche of a Rydberg gas both evolve to quasi-equilibria of electrons, ions and high-Rydberg neutrals \cite{Rolston_expand,Gallagher}.  

Early efforts to understand plasmas formed by Rydberg gas avalanche paid particular attention to the process of initiation.  Evolution to plasma in effusive atomic beams was long known for high-Rydberg gases of caesium and well explained by coupled rate equations \cite{Vitrant}.  But, low densities and ultracold velocity distributions were thought to exclude Rydberg-Rydberg collisional mechanisms in a MOT.  

In work on ultracold Rydberg gases of Rb and Cs, Gallagher, Pillet and coworkers describe the initial growth of electron signal by a model that includes ionization by blackbody radiation and collisions with a background of uncooled Rydberg atoms \cite{Mourachko,Gallagher,Li,Comparat,Tanner}. This picture was subsequently refined to include many-body excitation and autoionization, as well as attractive dipole-dipole interactions \cite{Viteau,Pillet}, later confirmed by experiments at Rice \cite{Mcquillen}.  

The Orsay group also studied the effect of adding Rydberg atoms to an established ultracold plasma.  They found that electron collisions in this environment completely ionize added atoms, even when selected to have deep binding energies \cite{Vanhaecke}.  They concluded from estimates of electron trapping efficiency that the addition of Rydberg atoms does not significantly alter the electron temperature of the plasma.  

Tuning pair distributions by varying the wavelength of the excitation laser, Weidem\"uller and coworkers confirmed the mechanical effects of van der Waals interactions on the rates of Penning ionization in ultracold $^{87}$Rb Rydberg gases \cite{Amthor_mech}.  They recognized blackbody radiation as a possible means of final-state redistribution, and extended this mechanical picture to include long-range repulsive interactions \cite{Amthor_model}.  This group later studied the effects of spatial correlations in the spontaneous avalanche of Rydberg gases in a regime of strong blockade, suggesting a persistence of initial spatial correlations \cite{RobertdeSaintVincent}.  

Robicheaux and coworkers have recently investigated the question of prompt many-body ionization from the point of view of Monte Carlo classical trajectory calculations \cite{Goforth}.  For atoms on a regular or random grid driven classically by an electromagnetic field, they find that many-body excitation enhances prompt ionization by about twenty percent for densities greater than $5.6 \times 10^{-3}/(n_0^2 a_0)^3$, where $n_0$ is the principal quantum number of the Rydberg gas and $a_0$ is the Bohr radius.  They observed that density fluctuations (sampled from the distribution of nearest neighbour distances) have a greater effect, and point to the possible additional influence of secondary electron-Rydberg collisions and the Penning production of fast atoms not considered by the model, but already observed by Raithel and coworkers \cite{Knuffman}.  

The Raithel group also found direct evidence for electron collisional $\ell$-mixing in a Rb MOT \cite{Dutta}, and used selective field ionization to monitor evolution to plasma on a microsecond timescale in ultracold $^{85}$Rb $65d$ Rydberg gases with densities as low as $10^8$ cm$^{-3}$ \cite{WalzFlannigan}.  Research by our group at UBC has observed very much the same dynamics in the relaxation of Xe Rydberg gases of similar density prepared in a molecular beam \cite{Hung2014}.  In both cases, the time evolution to avalanche is well-described by coupled rate equations (see below), assuming an initializing density of Penning electrons determined by Robicheaux's criterion \cite{Robicheaux05}, applied to an Erlang distribution of Rydberg-Rydberg nearest neighbours.  

Theoretical investigations of ultracold plasma physics have focused for the most part on the long- and short-time dynamics of plasmas formed by direct photoionization \cite{PhysRept,Lyon}.  In addition to studies mentioned above, key insights on the evolution dynamics of Rydberg gases have been provided by studies of Pohl and coworkers exploring the effects of ion correlations and recombination-reionization on the hydrodynamics of plasma expansion \cite{Pohl:2003,PPR}.  Further research has drawn upon molecular dynamics (MD) simulations to reformulate rate coefficients for the transitions driven by electron impact between highly excited Rydberg states \cite{PVS}, and describe an effect of strong coupling as it suppresses three-body recombination \cite{Bannasch:2011}.  MD simulations confirm the accuracy of coupled rate equation descriptions for systems with $\Gamma$ as large as 0.3.  Newer calculations suggest a strong connection between the order created by dipole blockade in Rydberg gases and the most favourable correlated distribution of ions in a corresponding strongly coupled ultracold plasma \cite{Bannasch:2013}.  

Tate and coworkers have studied ultracold plasma avalanche and expansion theoretically as well as experimentally.  Modelling observed expansion rates, they recently found that $^{85}$Rb atoms in a MOT form plasmas with effective initial electron temperatures determined by initial Rydberg density and the selected initial binding energy, to the extent that these parameters determine the fraction of the excited atoms that ionize by electron impact in the avalanche to plasma \cite{Forest}.  This group also returned to the question of added Rydberg atoms, and managed to identify a crossover in $n_0$, depending on the initial electron temperature, that determines whether added Rydberg atoms of a particular initial binding energy act to heat or cool the electron temperature \cite{Crockett}.   

Our group has focused on the plasma that evolves from a Rydberg gas under the low-temperature conditions of a skimmed, seeded supersonic molecular beam.  In work on nitric oxide starting in 2008 \cite{Morrison2008,Plasma_expan,Morrison_shock,PCCP}, we established an initial kinetics of electron impact avalanche ionization that conforms with coupled rate equation models \cite{Saquet2011,Saquet2012,Scaling,haenelCP} and agrees at early times with the properties of ultracold plasmas that evolve from ultracold atoms in a MOT.  We have also observed unique properties of the NO ultracold plasma owing to the fact that its Rydberg states dissociate \cite{Haenel2017}, and identified relaxation pathways that may give rise to quantum effects \cite{SousMBL,SousNJP}.  The remainder of this review focuses on the nitric oxide ultracold plasma and the unique characteristics conferred by its evolution from a Rydberg gas in a laser-crossed molecular beam.

\section{Avalanche to strong coupling in a molecular Rydberg gas}

\subsection{The molecular beam ultracold plasma compared with a MOT}

When formed with sufficient density, a Rydberg gas of principal quantum number $n_0>30$ undergoes a spontaneous avalanche to form an ultracold plasma \cite{Li,Morrison2008,RobertdeSaintVincent}.  Collisional rate processes combine with ambipolar hydrodynamics to govern the properties of the evolving plasma.  For a molecular Rydberg gas, neutral fragmentation, occurs in concert with electron-impact ionization, three-body recombination and electron-Rydberg inelastic scattering.  Neutral dissociation combined with radial expansion in a shaped distribution of charged particles, can give rise to striking effects of self-assembly and spatial correlation \cite{Schulz-Weiling2016,Haenel2017}.   

The formation of a molecular ultracold plasma requires the conditions of local temperature and density afforded by a high mach-number skimmed supersonic molecular beam.  Such a beam propagates at high velocity in the laboratory, with exceedingly well-defined hydrodynamic properties, including a propagation-distance-dependent density and sub-Kelvin temperature in the moving frame \cite{MSW_tutorial}.  The low-temperature gas in a supersonic molecular beam differs in three important ways from the atomic gas laser-cooled in a magneto-optical trap (MOT).

The milli-Kelvin temperature of the gas of ground-state NO molecules entrained in a beam substantially exceeds the sub-100 micro-Kelvin temperature of laser-cooled atoms in a MOT.  However, the evolution to plasma tends to erase this distinction, and the two further characteristics that distinguish a beam offer important advantages for ultracold plasma physics:  Charged-particle densities in a molecular beam can exceed those attainable in a MOT by orders of magnitude.  A great many different chemical substances can be seeded in a free-jet expansion, and the possibility this affords to form other molecular ultracold plasmas, introduces interesting and potentially important new degrees of freedom governing the dynamics of their evolution.

\subsection{Supersonic molecular beam temperature and particle density}

Seeded in a skimmed supersonic molecular beam, nitric oxide forms different phase-space distributions in the longitudinal (propagation) and transverse coordinate dimensions.  As it propagates in $z$, the NO molecules reach a terminal laboratory velocity, $u_{\parallel}$, of about 1400 ${\rm ms^{-1}}$, which varies with the precise seeding ratio.  

The distribution of $v_{\parallel}$, narrows to define a local temperature, $T_{\parallel}$, of approximately 0.5 K.  The beam forms a Gaussian spatial distribution in the transverse coordinates, $x$ and $y$.  In this plane, the local velocity, $v_{\perp}(r)$ is defined for any radial distance almost entirely by the divergence velocity of the beam, $u_{\perp}(r)$.  Phase-space sorting cools the temperature in the transverse coordinates, $T_{\perp}$ to a value as low as $\sim 5$ mK \cite{MSW_tutorial}.  

The stagnation pressure and seeding ratio determine the local density distribution as a function of $z$.  For example, expanding from a stagnation pressure of 500 kPa with a 1:10 seeding ratio, a molecular beam propagates 2.5 cm to a skimmer and then 7.5 cm to a point of laser interaction, where it contains NO at a peak density of $1.6 \times 10^{14}$ cm$^{-3}$.  

Here, crossing the molecular beam with a laser beam tuned to the transition sequence, ${\rm X} ~^2 \Pi_{1/2} ~N'' = 1 \xrightarrow{\omega_1} {\rm A} ~^2\Sigma^+ ~N'=0  \xrightarrow{\omega_2} n_0 f(2)$ forms a Gaussian ellipsoidal volume of Rydberg gas in a single selected principal quantum number, $n_0$, orbital angular momentum, $\ell = 3$, NO$^+$ core rotational quantum number, $N^+ = 2$ and total angular momentum neglecting spin, $N=1$.  

A typical $\omega_1$ pulse energy of 2 $\mu$J and a Gaussian width of 0.2 mm serves to drive the first step of this sequence in a regime of linear absorption.  Overlapping this volume by an $\omega_2$ pulse with sufficient fluence to saturate the second step forms a Rydberg gas ellipsoid with a nominal peak density of $5 \times 10^{12}$ cm$^{-3}$  \cite{Morrison2008,MSW_tutorial}.  Fluctuations in the pulse energy and longitudinal mode of $\omega_1$ cause the real density to vary.  For certain experiments, we find it convenient to saturate the $\omega_1$ transition, and vary the density of Rydberg gas by delaying $\omega_2$.  An $\omega_1$-$\omega_2$ delay, $\Delta t$, reduces the Rydberg gas density by a precise factor, $e^{-\Delta t/\tau}$, where $\tau$ is the 200 ns radiative lifetime of NO ${\rm A} ~^2\Sigma^+ ~N'=0$ \cite{Carter,Hancock}.

\subsection{Penning ionization}

The density distribution of a Rydberg gas defines a local mean nearest neighbour distance, or Wigner-Seitz radius of $ a_{ws} =  \left(3/4 \pi \rho \right)^{1/3} $, where $\rho$ refers to the local Rydberg gas density.  For example, a Rydberg gas with a density of $ \rho_0=0.5 \times 10^{12}$ cm$^{-3} $ forms an Erlang distribution \cite{Torquato.1990} of nearest neighbour separations with a mean value of $ 2 a_{ws}=1.6$  $\mu$m.  

A semi-classical model \cite{Robicheaux05} suggests that 90 percent of Rydberg molecule pairs separated by a critical distance, $ r_c = 1.8 \cdot 2 n_0^2 a_0 $ or less undergo Penning ionization within 800 Rydberg periods.  We can integrate the Erlang distribution from $ r=0 $ to the critical distance $r = r_c$ for a Rydberg gas of given $n_0$, to define the local density of Penning electrons ($ \rho_e$ at $t=0$) produced by this prompt interaction, for any given initial local density, $\rho_0$ by the expression:
\begin{equation}
\rho_e(\rho_0,n_0) = \frac{0.9}{2} \cdot 4 \pi \rho_0 ^2\int_0^{r_{c}} r^2 \mathrm{e}^{-\frac{4\pi}{3}\rho_0 r^3}\mathrm{d}r \quad.
\label{eqn:Erlang}
\end{equation}

Evaluating this definite integral yields an equation in closed form that predicts the Penning electron density for any particular initial Rydberg density and principal quantum number.
\begin{equation}
\rho_e(\rho_0,n_0) =\frac{0.9 \rho_0}{2}(1-\mathrm{e}^{-\frac{4\pi}{3}\rho_0 r_c^3}) \quad.
\label{Eq:PenDens}
\end{equation}
\begin{figure}[h!]
\centering
\includegraphics[scale=0.33]{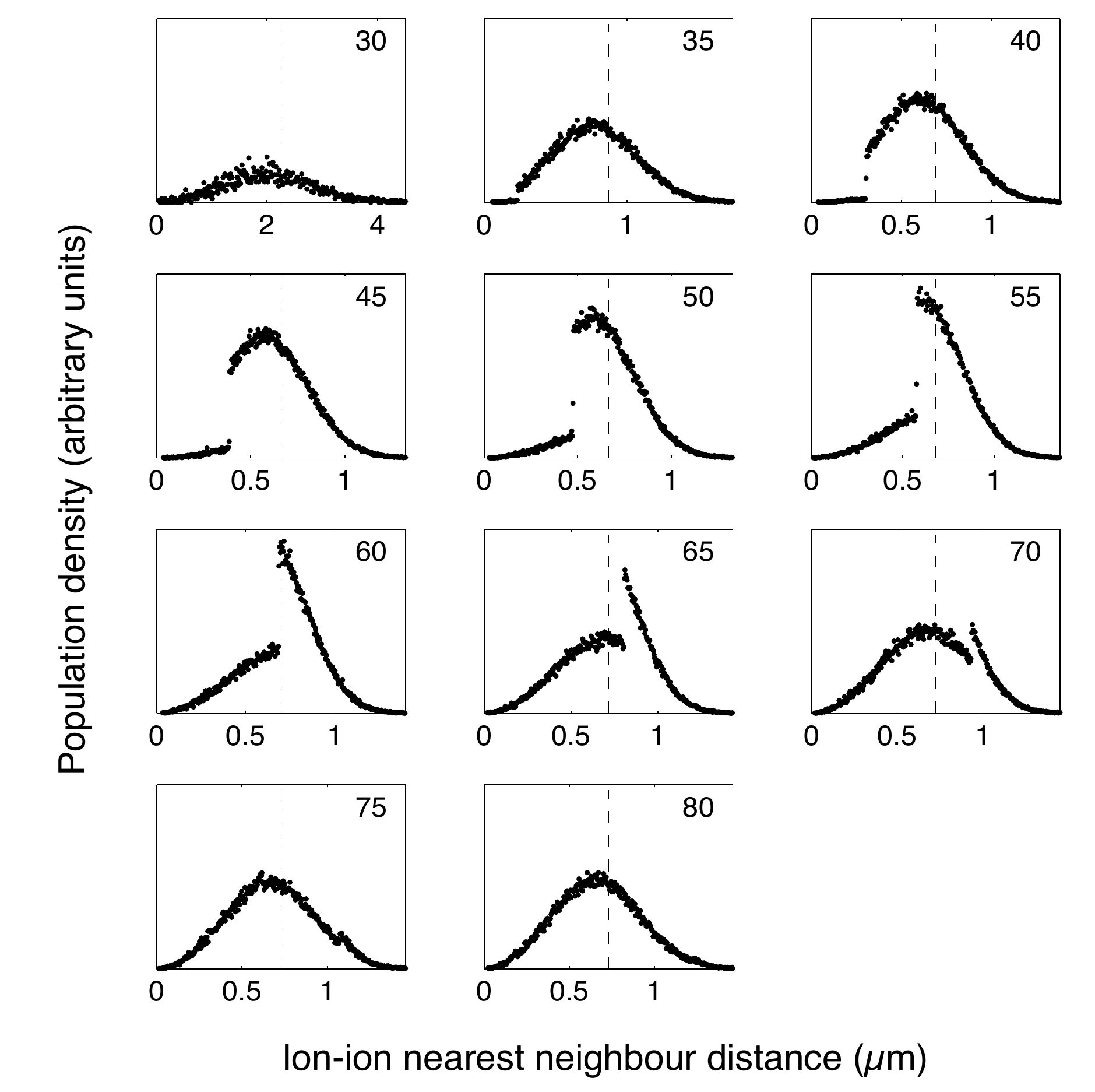}
\caption{Distributions of ion-ion nearest neighbours following Penning ionization and electron-impact avalanche simulated for a predissociating molecular Rydberg gas of initial principal quantum number, $n_0$, from 30 to 80, and density of 10$^{12}$ cm$^{-3}$.  Dashed lines mark corresponding values of $a_{ws}$. Calculated by counting ion distances after relaxation to plasma in 10$^6$-particle stochastic simulations. Integrated areas proportional to populations surviving neutral dissociation.}
\label{fig:PL}
\end{figure}

Prompt Penning ionization acts on the portion of the initial nearest-neighbour distribution in the Rydberg gas that lies within $r_c$.  When a molecule ionizes, its collision partner relaxes to a lower principal quantum number, $n'<n_0/\sqrt{2}$.  This close-coupled interaction disrupts the separability of Rydberg orbital configurations in the Penning partner.  This causes mixing with core penetrating states that are strongly dissociative.  Penning partners are thus very likely to dissociate, leaving a spatially isolated distribution of ions.  We refer to the spatial correlation that results as a Penning lattice \cite{Sadeghi:2014}.  The extent of this effect varies depending on the local density and the selected initial principal quantum number.  Figure \ref{fig:PL} shows the degree to which Rydberg gases with initial principal quantum numbers from 30 to 80 form a Penning lattice for an initial density of $1 \times 10^{12} ~{\rm cm}^{-3}$.  

\subsection{Spontaneous electron-impact avalanche}

The electrons produced by prompt Penning ionization start an electron impact avalanche.  The kinetics of this process are well described by a set of coupled rate equations that account for state-to-state electron-Rydberg inelastic scattering, electron-impact ionization and three-body ion-electron recombination \cite{PPR,Saquet2011,Saquet2012,Scaling} using detailed rate coefficients,  $k_{ij}$, $k_{i,ion}$ and $k_{i,tbr}$ validated by MD simulations \cite{PVS}.  
\begin{eqnarray}
-\frac{d\rho_i}{dt}&=&\sum_{j}{k_{ij}\rho_e\rho_i}-\sum_{j}{k_{ji}\rho_e\rho_j} \nonumber\\
&& +k_{i,ion}\rho_e\rho_i-k_{i,tbr}\rho^3_e 
  \label{level_i}
\end{eqnarray}
\noindent and,
\begin{equation}
\frac{d\rho_e}{dt}=\sum_{i}{k_{i,ion}\rho_e^2}-\sum_{i}{k_{i,tbr}\rho^3_e}
  \label{electron}
\end{equation}

The relaxation of Rydberg molecules balances with collisional ionization to determine an evolving temperature of avalanche electrons to conserve total energy per unit volume. 
\begin{equation}
E_{tot}=\frac{3}{2}k_BT_e(t)\rho_e(t)-R\sum_i{\frac{\rho_i(t)}{n_i^2}},
  \label{energy}
\end{equation}
Here, for simplicity, we neglect the longer-time effects of Rydberg predissociation and electron-ion dissociative recombination \cite{Saquet2012}.

Such calculations show that the conversion from Rydberg gas to plasma occurs on a timescale determined largely by the local Penning electron density, or Penning fraction, $P_f = \rho_e/\rho_0$, which depends on the local density of Rydberg molecules and their initial principal quantum number.  

Avalanche times predicted by coupled rate equation calculations range widely.  For example, in a model developed for experiments on xenon, simulations predict that a Rydberg gas with $n_0 = 42$ at a density of $8.8 \times 10^8 ~{\rm cm}^{-3}$ ($P_f = 6 \times 10^{-5}$) avalanches with a half time of  40 $\mu$s \cite{Hung2014}.  At an opposite extreme, rate equations estimate that a Rydberg gas of NO with $n_0=60$ at a density of $1 \times 10^{12} ~{\rm cm}^{-3}$ ($P_f = 0.3$) rises to plasma in about 2 ns \cite{Saquet2012}.  

\begin{figure}[h!]
\centering
\includegraphics[width= .49 \textwidth]{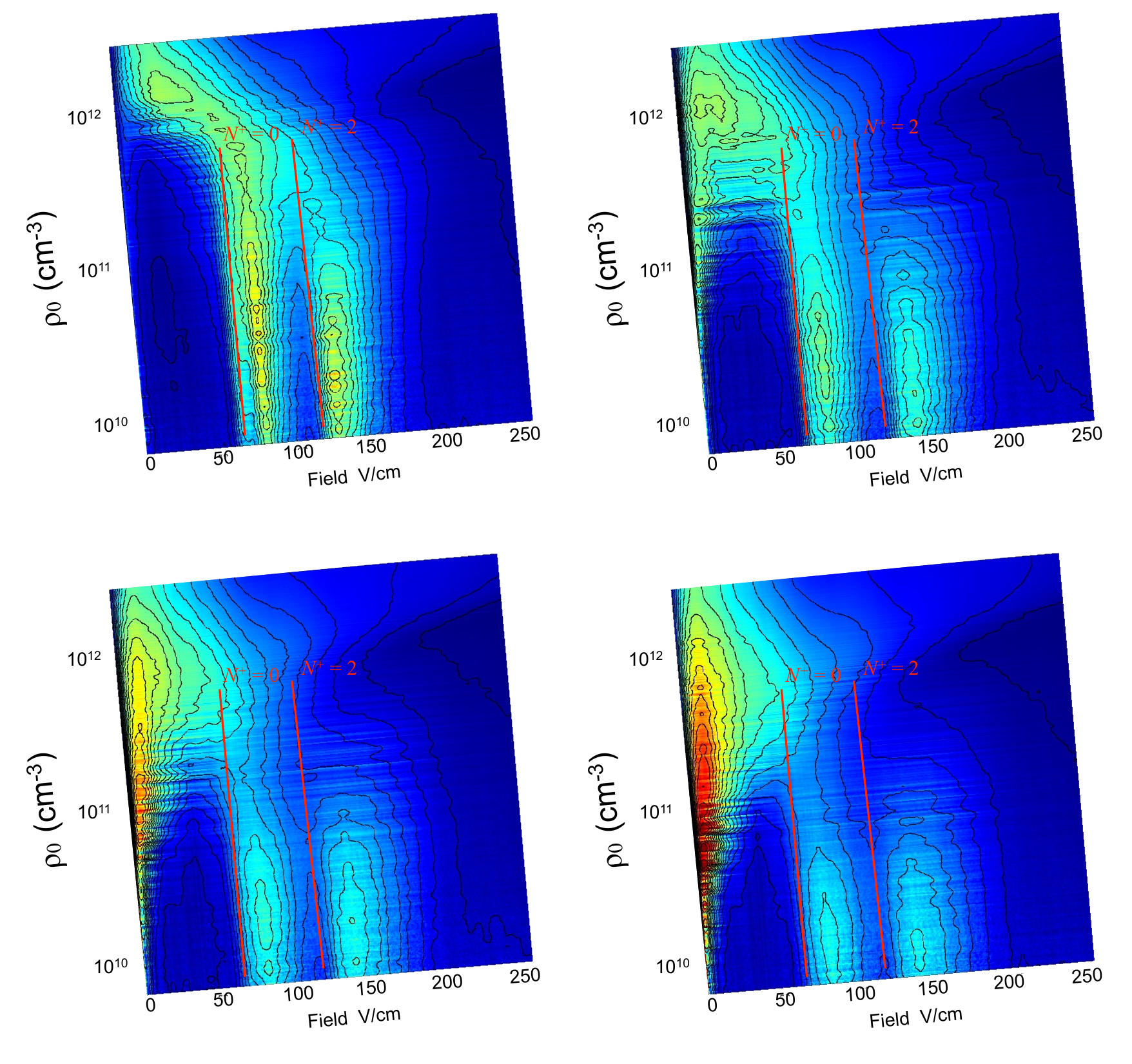}
   \caption{Contour plots showing SFI signal as a function the applied field for an $nf(2)$ Rydberg gas with an initial principal quantum number, $n_0=49$.  Each frame represents 4,000 SFI traces, sorted by initial Rydberg gas density.  Ramp field beginning at 0 and 150 ns (top, left to right), and  300 and 450 ns (bottom) after the $\omega_2$ laser pulse.  The two bars of signal most evident at early ramp field delay times represent the field ionization of the $49f(2)$ Rydberg state respectively to NO$^+$ X $^1\Sigma^+$ cation rotational states, $N^+=0$ and 2.  The signal waveform extracted near zero applied field represents the growing population of plasma electrons.  
   }
\label{fig:SFI}
\end{figure}

Selective field ionization (SFI) probes the spectrum of binding energies in a Rydberg gas.  Applied as a function of time after photoexcitation, SFI maps the evolution from a state of selected initial principal quantum number, $n_0$, to plasma \cite{Haenel2017}.  Figure \ref{fig:SFI} shows SFI spectra taken at a sequence of delays after the formation of $49f(2)$ Rydberg gases of varying density.    

Here, we can see that a $49f(2)$ Rydberg gas with an estimated initial density $\rho_0 = 3 \times 10^{11} ~{\rm cm}^{-3}$ relaxes to plasma on a timescale of about 500 ns.  Observations such as these agree well with the predictions of coupled rate-equation calculations.  We can understand this variation in relaxation dynamics with $\rho_0$ and $n_0$ quite simply in terms of the corresponding density of prompt Penning electrons these conditions afford to initiate the avalanche to plasma.  

\begin{figure}[h!]
\centering
\includegraphics[width= .4 \textwidth]{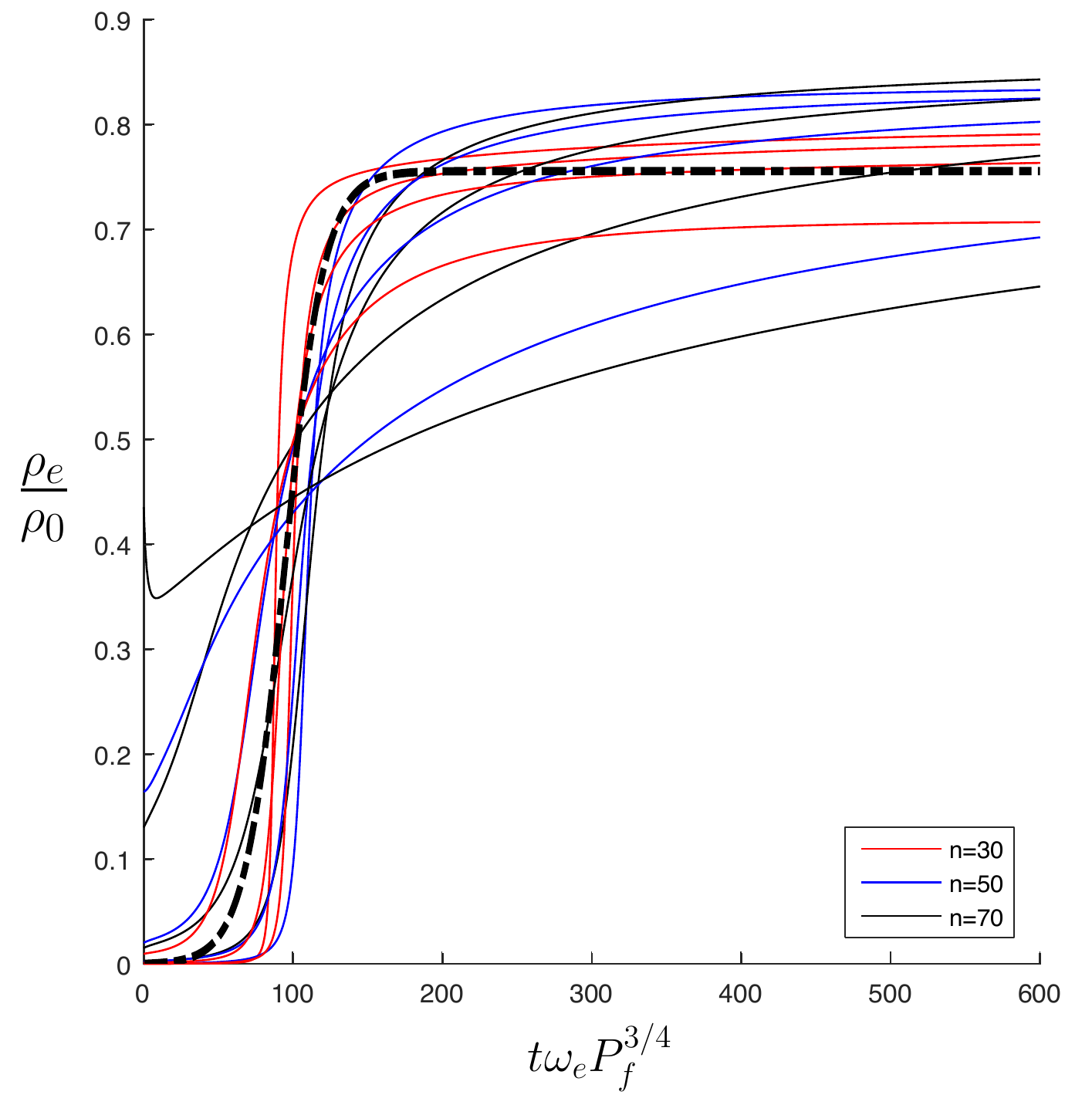}
   \caption{Rise in fractional electron density as a function of time scaled by the plasma frequency, $\omega_e$ and fraction, $\rho_e(t=0)/\rho_0 = P_f$, of prompt Penning electrons.  Simulation results shown for $n_0 = 30$, 50 and 70 with initial densities, $\rho_0 = 10^9,~10^{10},~10^{11},~{\rm and}~10^{12}~{\rm cm}^{-3}$.  
   }
\label{fig:scaled_rise}
\end{figure}

Figure \ref{fig:scaled_rise} illustrates this, showing how rise times predicted by coupled rate-equation simulations for a large range of initial densities and principal quantum number match when plotted as a function of time scaled by the ultimate plasma frequency and fraction of prompt Penning electrons.  The dashed line gives an approximate account of the scaled rate of avalanche under all conditions of Rydberg gas density and initial principal quantum number in terms of the simple sigmoidal function:

\begin{equation}
\frac{\rho_e}{\rho_0} = \frac{a}{b+e^{-c\tau}},
  \label{scaledEq1}
\end{equation}
where,
\begin{equation}
\tau = t \omega_e P_f^{3/4},
  \label{scaledEq2}
\end{equation}
in which $\omega_e$ is the plasma frequency after avalanche, $P_f$ is the fraction of prompt Penning electrons, and $a = 0.00062$,  $b =   0.00082$ and $c =     0.075$ are empirical coefficients.  

\subsection{Evolution to plasma in a Rydberg gas Gaussian ellipsoid}

As outlined above, the local density and principal quantum number together determine the rate at which a Rydberg gas avalanches to plasma.  Our experiment crosses a 2 mm wide cylindrically Gaussian molecular beam with a 1 mm diameter TEM$_{00}$ $\omega_1$ laser beam to produce a Gaussian ellipsoidal distribution of molecules excited to the A $^2\Sigma^+$ $v=0, ~N'=0$ intermediate state.  A larger diameter $\omega_2$ pulse then drives a second step that forms a Rydberg gas in a single $n_0f(2)$ state with the spatial distribution of the intermediate state.  

We model this shaped Rydberg gas as a system of 100 concentric ellipsoidal shells of varying density \cite{haenelCP}.  Coupled rate equations within each shell describe the avalanche to plasma.  This rate process proceeds from shell to shell with successively longer induction periods, determined by the local density as detailed above.  The rising conversion of Rydberg molecules to ions plus neutral dissociation products conserves the particle number in each shell.  We assume that local space charge confines electrons to shells, conserving quasi-neutrality.  Electrons exchange kinetic energy at the boundaries of each shell, which determines a single plasma electron temperature.  

\begin{figure}[h!]
\centering
\includegraphics[width= .5 \textwidth]{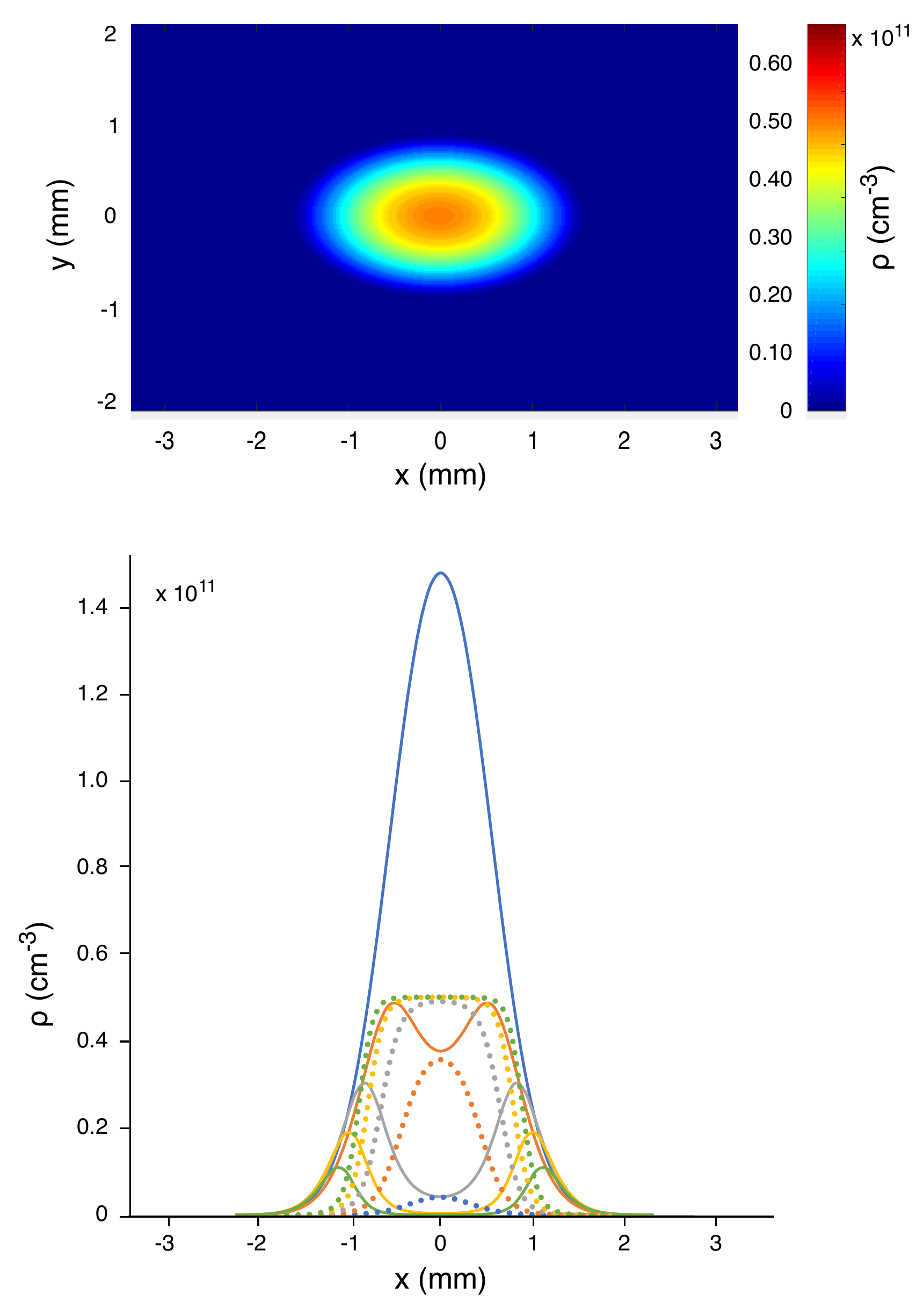}
   \caption{(top frame) Cross-sectional contour diagram in the $x,y$ plane for $z=0$ describing the distribution of ion plus electron density over 100 shells of Gaussian ellipsoid with initial dimensions, $\sigma_x= 0.75$ mm and $\sigma_y= \sigma_z = 0.42$ mm and an initial $n_0 = 50$ Rydberg gas density, $\rho_0 = 2 \times 10^{11}$ cm$^{3}$ after an evolution time of 100 ns.  (bottom frame) Curves describing the (dashed) ascending ion and (solid) descending Rydberg gas densities of each shell as functions of evolution time, for $t=20$, 40, 60, 80 and 100 ns.  
   }
\label{fig:shell}
\end{figure}

The upper frame of Figure \ref{fig:shell} shows contours of NO$^+$ ion density after 100 ns obtained from a shell-model coupled rate-equation simulation of the avalanche of a Gaussian ellipsoidal Rydberg gas of nitric oxide with a selected initial state, $50f(2)$ and a density of $2 \times 10^{11}$ cm$^{-3}$.  Here, we simulate a relaxation that includes channels of predissociation at every Rydberg level and redistributes the energy released to electrons, which determines a uniform rising electron temperature for all shells.  

For comparison, the lower frame plots curves describing the ion density of each shell as a function of time from 20 to 100 ns, as determined by applying Eq \ref{scaledEq1} for the local conditions of initial Rydberg gas density.  This numerical approximation contains no provision for predissociation.  Coupled rate-equation simulations for uniform volumes show that predissociation depresses yield to some degree, but has less effect on the avalanche kinetics \cite{Saquet2012}.  Therefore, we can expect sets of numerically estimated shell densities, scaled to agree with the simulated ion density at the elapsed time of 100 ns to provide a reasonable account of the earlier NO$^+$ density profiles as a function of time.

\begin{figure}[h!]
\centering
\includegraphics[width= .4 \textwidth]{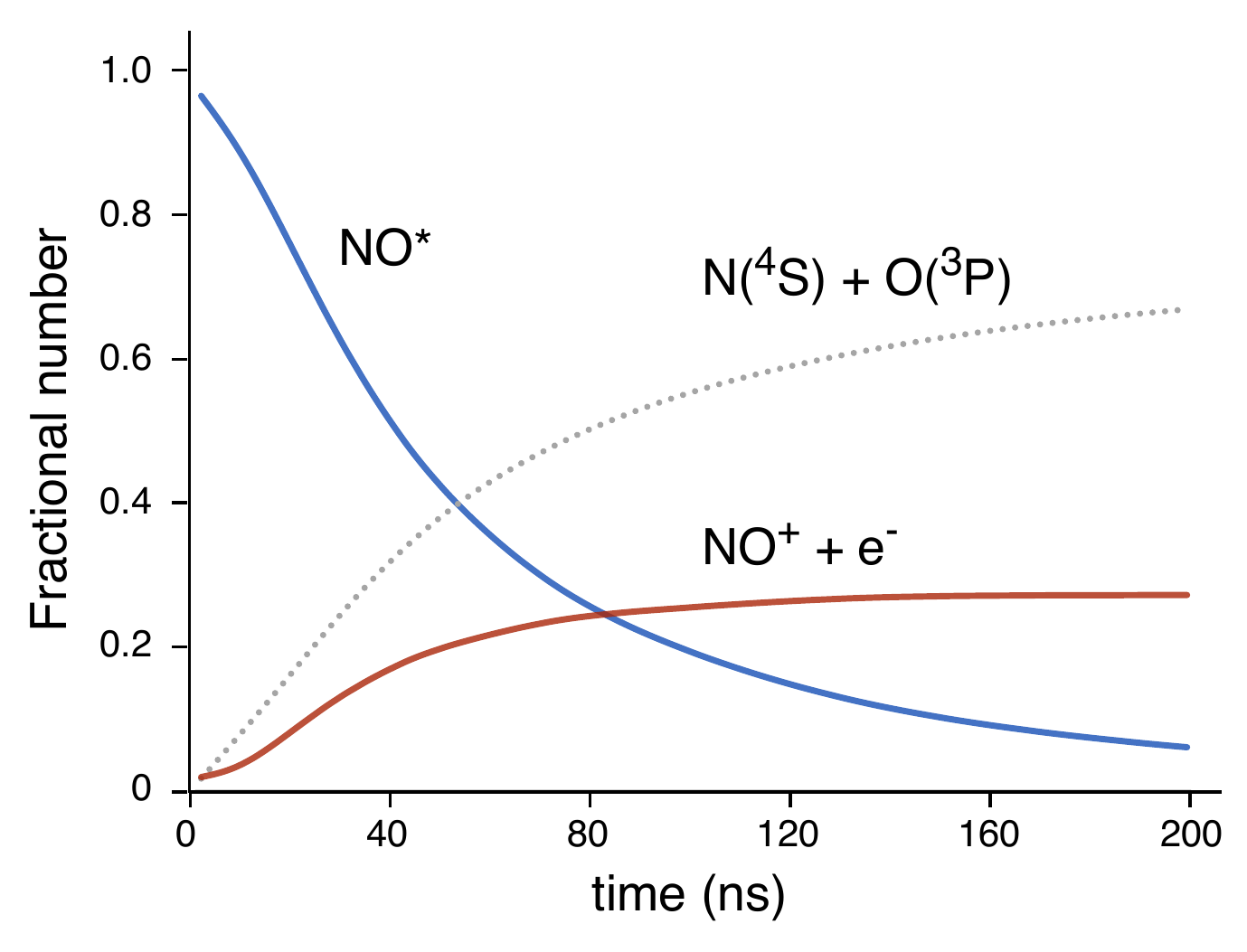}
   \caption{Global population fractions of particles as they evolve in the avalanche of a shell-model ellipsoidal Rydberg gas with the initial principal quantum number and density distribution of Figure \ref{fig:shell}
   }
\label{fig:shell_yields}
\end{figure}

For each time step, the difference, $\rho_0 - \rho_e$ defines the neutral population of each shell.  We assign a fraction of this population to surviving Rydberg molecules, such that the total population of NO$^*$ as a function of time agrees with the prediction of the shell-model simulation, as shown in Figure \ref{fig:shell_yields}.  We consider the balance of this neutral population to reflect NO$^*$ molecules that have dissociated to form N($^4$S) $+$ O($^3$P).  Figure \ref{fig:shell} plots these surviving Rydberg densities as a function of radial distance for each evolution time.  At the initial density of this simulation, note at each time step that a higher density of Rydberg molecules encloses the tail of the ion density distribution in $x$.   

\subsection{Plasma expansion and NO$^+$ - NO$^*$ charge exchange as an avenue of quench}

We regard the ions as initially stationary.  The release of electrons creates a radial electric potential gradient, which gives rise to a force, $-e\nabla \phi_{k,j}(t)$, that accelerates the ions in shell $j$ in direction $k$ according to \cite{Sadeghi.2012}:

\begin{align}
\frac{-e}{m'}\nabla \phi_{k,j}(t) = & \frac{\partial u_{k,j}(t)}{\partial t} \notag  \\
= & \frac{k_BT_e(t)}{m'\rho_j(t)} \frac{\rho_{j+1}(t) - \rho_j(t)}{r_{k,j+1}(t) - r_{k,j}(t)},
  \label{dr_dt}
\end{align}

\noindent where $\rho_j(t)$ represents the density of ions in shell $j$.  

The instantaneous velocity, $u_{k,j}(t)$ determines the change in the radial coordinates of each shell, $r_{k,j}(t)$, 
\begin{equation}
\frac{\partial r_{k,j}(t)}{\partial t}=u_{k,j}(t) = \gamma_{k,j}(t) r_{k,j}(t),
  \label{dr_dt}
\end{equation}
\noindent which in turn determines shell volume and thus its density, $ \rho_j(t)$.   
The electron temperature supplies the thermal energy that drives this ambipolar expansion.  Ions accelerate and $T_e$ falls according to: 

\begin{equation}
\frac{3k_B}{2}\frac{\partial T_e(t)}{\partial t}= -\frac{m'}{\sum_{j}{N_j}}\sum_{k,j}{N_j u_{k,j}(t)\frac{\partial u_{k,j}(t)}{\partial t}},
  \label{dr_dt}
\end{equation}
\noindent where we define an effective ion mass, $m'$, that recognizes the redistribution of the electron expansion force over all the NO$^+$ charge centres by resonant ion-Rydberg charge exchange, which occurs with a very large cross section \cite{PPR}.  
\begin{equation}
m' =\left (1+ \frac{\rho^*_{j}(t)}{ \rho_j(t)}\right) m ,
  \label{dr_dt}
\end{equation}
\noindent in which $\rho^*_{j}(t)$ represents the instantaneous Rydberg density in shell $j$.

The initial avalanche in the high-density core of the ellipsoid leaves few Rydberg molecules, so this term has little initial effect.  Rydberg molecules predominate in the lower-density wings.  There, momentum sharing by charge exchange assumes a greater importance.  

We see this most directly in the $\omega_2$ absorption spectrum of transitions to states in the $n_0 f(2)$ Rydberg series, detected as the long-lived signal that survives a flight time of 400 $\mu$s to reach the imaging detector.  The balance between the rising density of ions and the falling density of Rydberg molecules depends on the initial density of electrons produced by prompt Penning ionization.  As clear from Eq \ref{Eq:PenDens}, this Penning fraction depends sensitively on the principal quantum number, and for all principal quantum numbers, on the initial Rydberg gas density.  

\begin{figure}[h!]
\centering
\includegraphics[width= .4 \textwidth]{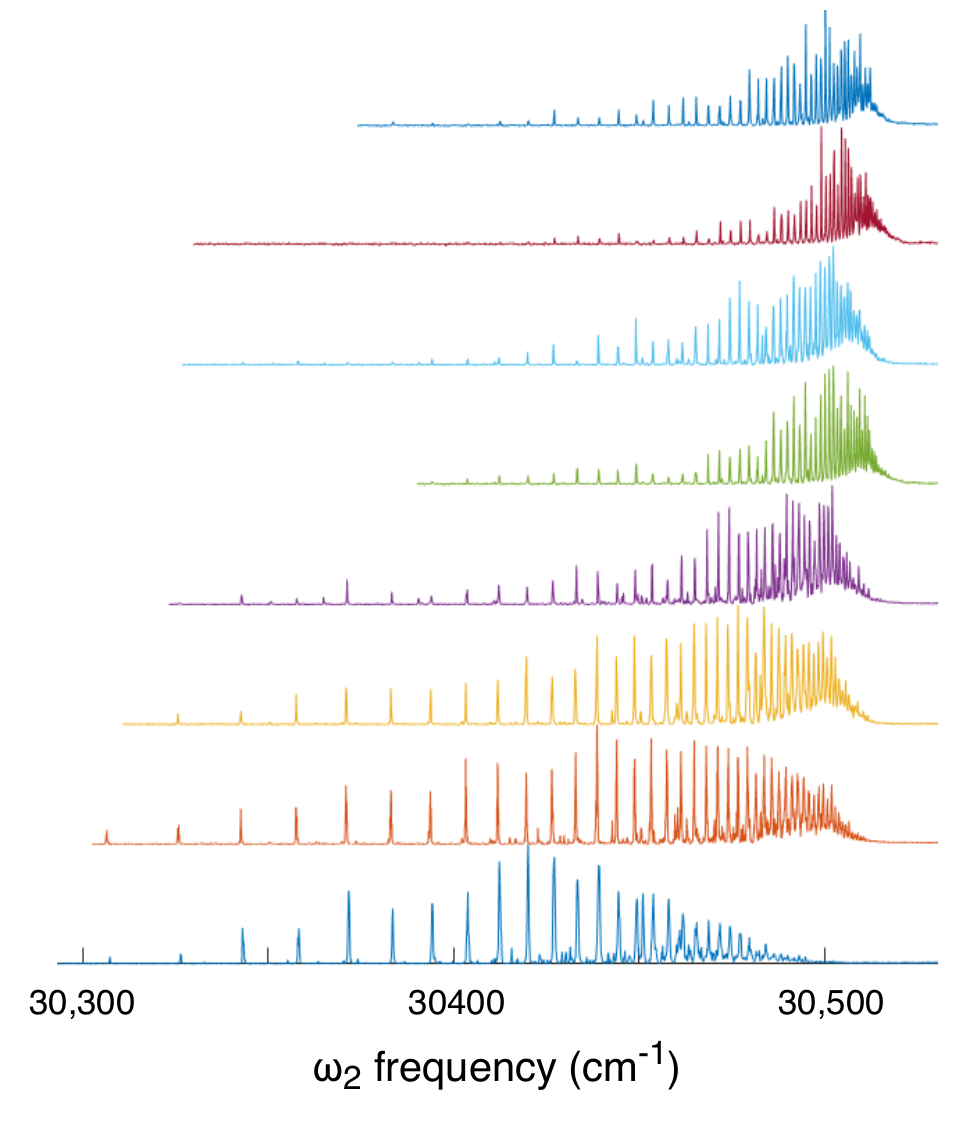}
   \caption{Double-resonant spectra of nitric oxide Rydberg states in the $nf$ series converging to NO$^+$ $v=0$, $N^+=2$ (designated, $nf(2)$), derived from the late-peak signal obtained after a flight time of 400 $\mu$s by scanning $\omega_2$ for $\omega_1$ tuned to NO A $^2\Sigma^+$ $v=0$, $N'=0$ for initial $nf(2)$ densities from top to bottom of 0.07, 0.10, 0.13, 0.19, 0.27, 0.30, 0.32 and $3 \times 10^{12}$ cm$^{3}$.  
  }
\label{fig:w2_spectra}
\end{figure}

Figure \ref{fig:w2_spectra} shows a series of $\omega_2$ late-signal excitation spectra for a set of initial densities.  Here, we see a clear consequence of the higher-order dependence of Penning fraction - and thus the NO$^+$ ion - NO$^*$ Rydberg molecule balance - on $n_0$, the $\omega_2$-selected Rydberg gas initial principal quantum number.  This Penning-regulated NO$^+$ ion - NO$^*$ Rydberg molecule balance appears necessary as a critical factor in achieving the long ultracold plasma lifetime required to produce this signal.  We are progressing in theoretical work that explains the stability apparently conferred by this balance.

\subsection{Bifurcation and arrested relaxation}

Ambipolar expansion quenches electron kinetic energy as the initially formed plasma expands.  Core ions follow electrons into the wings of the Rydberg gas.  There, recurring charge exchange between NO$^+$ ions and NO$^*$ Rydberg molecules redistributes the ambipolar force of the expanding electron gas, equalizing ion and Rydberg velocities.  This momentum matching effectively channels electron energy through ion motion into the overall $\pm x$ motion of gas volumes in the laboratory.  The internal kinetic energy of the plasma, which at this point is defined almost entirely by the ion-Rydberg relative motion, falls.  Spatial correlation develops, and over a period of 500 ns, the system forms the plasma/high-Rydberg quasi-equilibrium dramatically evidenced by the SFI results in Figure \ref{fig:SFI}.  

\begin{figure}[h!]
\centering
\includegraphics[width= .4 \textwidth]{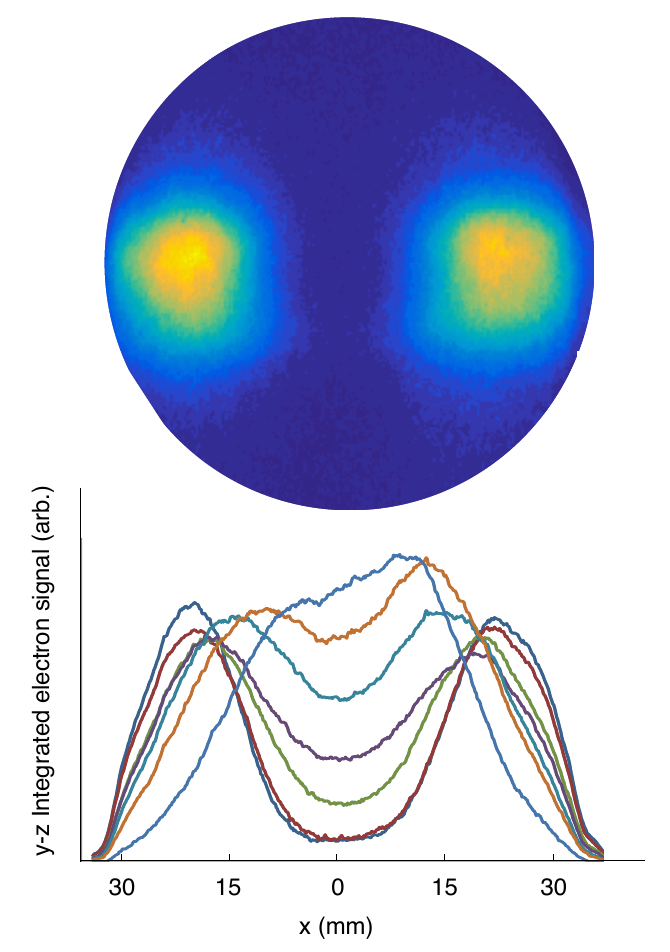}
   \caption{$x,y$ detector images of ultracold plasma volumes produced by 2:1 aspect ratio ellipsoidal Rydberg gases with selected initial state, $40f(2)$ after a flight time of 402 $\mu$s over a distance of 575 mm.  Lower frame displays the distribution in $x$ of the charge integrated in $y$ and $z$.  Both images represent the unadjusted raw signal acquired in each case after 250 shots.  
   }
\label{fig:bifurcation}
\end{figure}

In the wings, momentum redistribution owing to cycles of ion-Rydberg charge transfer retards radial expansion \cite{Pohl2003,PPR}.  By redirecting electron energy from ambipolar acceleration to $\pm x$ plasma motion, NO$^+$ to NO$^*$ charge exchange dissipates electron thermal energy.  This redistribution of energy released in the avalanche of the Rydberg gas to plasma, causes the ellipsoidal Rydberg gas to bifurcate \cite{Schulz-Weiling2016,Haenel2017}, forming very long-lived, separating charged-particle distributions.  We capture the electron signal from these recoiling volumes on an imaging detector as pictured in Figure \ref{fig:bifurcation}.  Here, momentum matching preserves density and enables ions and Rydberg molecules to relax to positions that minimize potential energy, building spatial correlation.  

The semi-classical description of avalanche and relaxation outlined above forms an important point of reference from which to interpret our experimental observations.  The laser crossed molecular beam illumination geometry creates a Rydberg gas with a distinctively shaped high-density spatial distribution.  This initial condition has an evident effect on the evolution dynamics.  We have developed semi-classical models that explicitly consider the coupled rate and hydrodynamic processes governing the evolution from Rydberg gas to plasma using a realistic, ellipsoidal representation of the ion/electron and Rydberg densities \cite{haenelCP}.  No combination of initial conditions can produce a simulation that conforms classically with the state of arrested relaxation we observe experimentally.

\subsection{A molecular ultracold plasma state of arrested relaxation}

Thus, we find that spontaneous avalanche to plasma splits the core of an ellipsoidal Rydberg gas of nitric oxide. As ambipolar expansion quenches the electron temperature of this core plasma, long-range, resonant charge transfer from ballistic ions to frozen Rydberg molecules in the wings of the ellipsoid quenches the ion-Rydberg molecule relative velocity distribution. This sequence of steps gives rise to a remarkable mechanics of self-assembly, in which the kinetic energy of initially formed hot electrons and ions drives an observed separation of plasma volumes. These dynamics redistribute ion momentum, efficiently channeling electron energy into a reservoir of mass-transport. This starts a process that evidently anneals separating volumes to a state of cold, correlated ions, electrons and Rydberg molecules. 

We have devised a three-dimensional spin model to describe this arrested state of the ultracold plasma in terms of two, three and four-level dipole-dipole energy transfer interactions (spin flip-flops), together with Ising interactions that arise from the concerted pairwise coupling of resonant pairs of dipoles \cite{SousMBL,SousNJP}.  

The Hamiltonian includes the effects of onsite disorder owing to the broad spectrum of states populated in the ensemble and the unique electrostatic environment of every dipole.  Extending ideas developed for simpler systems \cite{Burin1,Sondhi}, one can make a case for slow dynamics, including an arrest in the relaxation of NO Rydberg molecules to predissociating states of lower principal quantum number.

Systems of higher dimension ought to thermalize by energy transfer that spreads from rare but inevitable ergodic volumes (Griffiths regions) \cite{Sarang2, Roeck_griffith, RareRegions_rev, Thermal_inclusions}.  However, a feature in the self-assembly of the molecular ultracold plasma may preclude destabilization by rare thermal domains:  Whenever the quenched plasma develops a delocalizing Griffiths region, the local predissociation of relaxing NO molecules promptly proceeds to deplete that region to a void of no consequence.

In summary, the classical dynamics of avalanche and bifurcation appear to create a quenched condition of low temperature and high disorder in which dipole-dipole interactions drive self-assembly to a localized state purified by the predissociation of thermal regions.  We suggest that this state of the quenched ultracold plasma offers an experimental platform for studying quantum many-body physics of disordered systems. \\

\section{Acknowledgments}
This work was supported by the US Air Force Office of Scientific Research (Grant No. FA9550-17-1-0343), together with the Natural Sciences and Engineering research Council of Canada (NSERC), the Canada Foundation for Innovation (CFI) and the British Columbia Knowledge Development Fund (BCKDF). 

\bibliography{UCP_review.bib,Intro_ref}

\end{document}